\documentclass[sigconf]{acmart}

\AtBeginDocument{%
  \providecommand\BibTeX{{%
    \normalfont B\kern-0.5em{\scshape i\kern-0.25em b}\kern-0.8em\TeX}}}

\setcopyright{rightsretained}
\copyrightyear{2021}
\acmYear{2021}
\setcopyright{rightsretained}
\acmConference[UIST '21]{The 34th Annual ACM Symposium on User Interface Software and Technology}{October 10--14, 2021}{Virtual Event, USA}
\acmBooktitle{The 34th Annual ACM Symposium on User Interface Software and Technology (UIST '21), October 10--14, 2021, Virtual Event, USA}\acmDOI{10.1145/3472749.3474765}
\acmISBN{978-1-4503-8635-7/21/10}

\usepackage[colorinlistoftodos]{todonotes}
\usepackage{multicol}
\usepackage{caption}
\usepackage{subcaption}
\usepackage{soul}

\newcommand{\bw}[1]{{\color{black}#1}}

\begin{document}

\title{Screen2Words: Automatic Mobile UI Summarization with Multimodal Learning}

\author{Bryan Wang}
\authornote{This work was completed while the author was an intern at Google Research.}
\affiliation{%
  \institution{University of Toronto}
  \city{Toronto}
  \state{ON}
  \country{Canada}}
\email{bryanw@dgp.toronto.edu} 

\author{Gang Li}
\affiliation{%
  \institution{Google Research}
  \city{Mountain View}
  \state{CA}
  \country{USA}}
\email{leebird@google.com} 

\author{Xin Zhou}
\affiliation{%
  \institution{Google Research}
  \city{Mountain View}
  \state{CA}
  \country{USA}}
\email{zhouxin@google.com} 

\author{Zhourong Chen}
\affiliation{%
  \institution{Google Research}
  \city{Mountain View}
  \state{CA}
  \country{USA}}
\email{zrchen@google.com} 

\author{Tovi Grossman}
\affiliation{%
  \institution{University of Toronto}
  \city{Toronto}
  \state{ON}
  \country{Canada}}
\email{tovi@dgp.toronto.edu}

\author{Yang Li}
\affiliation{%
  \institution{Google Research}
  \city{Mountain View}
  \state{CA}
  \country{USA}}
\email{liyang@google.com}


\begin{abstract}
Mobile User Interface Summarization generates succinct language descriptions of mobile screens for conveying important contents and functionalities of the screen, which can be useful for many language-based application scenarios. We present Screen2Words, a novel screen summarization approach that automatically encapsulates essential information of a UI screen into a coherent language phrase. Summarizing mobile screens requires a holistic understanding of the multi-modal data of mobile UIs, including text, image, structures as well as UI semantics, motivating our multi-modal learning approach. We collected and analyzed a large-scale screen summarization dataset annotated by human workers. Our dataset contains more than 112k language summarization across $\sim$22k unique UI screens. We then experimented with a set of deep models with different configurations. Our evaluation of these models with both automatic accuracy metrics and human rating shows that our approach can generate high-quality summaries for mobile screens. We demonstrate potential use cases of Screen2Words and open-source our dataset and model to lay the foundations for further bridging language and user interfaces.

\end{abstract}




\begin{teaserfigure}
  \hspace*{0.45cm}   
  \centering
  \includegraphics[width=0.95\textwidth]{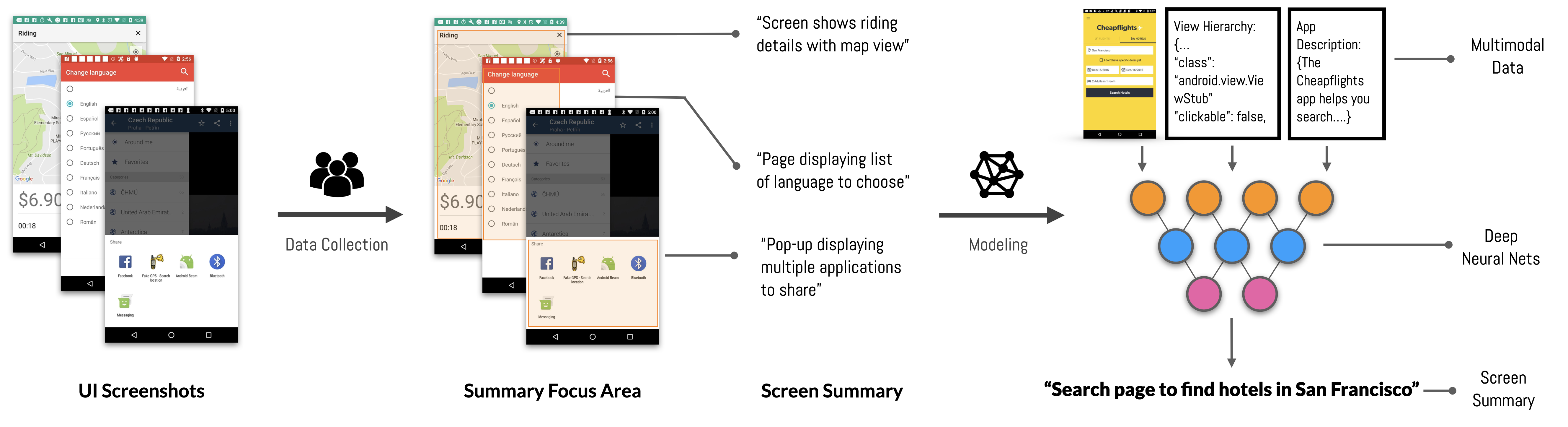}
  \caption{Screen2Words is a novel approach to automatically encapsulates essential information of a UI screen into a coherent language phrase. We collected the first large-scale screen summarization dataset, consisting of human annotations for 22,417 Android UI screens. We developed a set of deep learning models based on the dataset, leveraging the multi-modal data that a mobile screen carries. Our evaluation shows that our approach outperforms the heuristics baseline and is able to generate accurate summaries. }
  \Description{}
  \label{fig:teaser}
\end{teaserfigure}
\maketitle

\section{Introduction}
A mobile user interface screen often contains a rich set of graphical components from which a user can fulfill one or multiple functionalities. A succinct language description about a UI screen is useful for many language-based application scenarios. For example, a screen summary allows both conversational agents and end users to easily grasp the purpose and the state of a screen when accomplishing mobile tasks via language \cite{Toby2018, Toby2019}. A screen summary can be announced to help screen reader users quickly establish a mental model of an unknown mobile UI without waiting for the screen readers to scan through each element. \cite{ChallengesScreenReader_Rodrigues, Kuber2012DETERMININGTA}. More broadly, representing a UI screen in language, which is highly flexible and versatile, opens many opportunities for combining the strengths of the two communication mediums. However, screen summaries are largely nonexistent in existing applications. It is difficult to compose these summaries manually because user interfaces are highly diverse and dynamic. As such, we focus on automatic screen summarization, a task that generates compact yet informative language representation from semantic understanding of UI screens. 

Automatic screen summarization, while similar to image captioning~\cite{pmlr-v37-xuc15} and text summarization~\cite{text_summary_survey}, is unique in that it requires a holistic understanding of the multi-modal\footnote{The term multi-modal in this context refers to mechanisms that encode information from different data sources, which is different from the common HCI definition of the user interacting with devices using multiple modalities.} data that a mobile screen carries. Existing work has investigated using deep learning to encode multi-modal UI data into vector representations \cite{li2021screen2vec}, which has been shown useful for various downstream tasks such as UI retrieval \cite{swire2019, bunian2021vins}. However, few efforts have been made to bridge the gap between the learned semantic representation and natural language that can be communicated with human users. On the other hand, recent work \cite{li2020widget, zhang2021screen} predicts alt-text labels for icons and widgets on mobile screens by considering multi-modal data sources, yet they focus on individual components and do not generate phrases that can describe the entire mobile UI screen, which is a more challenging task.

In this paper, we present \textit{Screen2Words}, a novel generative approach for encapsulating complex information presented in a mobile UI into a succinct language description. To aid the development of Screen2Words, we collected the first large-scale screen summarization dataset, which consists of human annotations for 22,417 Android UI screens sampled \bw{from RICO \cite{Rico}, a public mobile screen corpus}. Inspired by recent work on representing GUI semantics using deep learning \cite{swire2019, li2021screen2vec, bunian2021vins, li2020widget}, we experimented with a set of deep models on the Screen2Words dataset to investigate the feasibility and effectiveness of our approach for screen summarization. We evaluated these models based on a set of accuracy metrics commonly used in image captioning and machine translation tasks. The result showed that all the deep models outperformed heuristic-based methods. 
A comparison between model variants showed that using multi-modal data sources about a UI screen leads to superior summarization accuracy. Our best results were achieved by using a combination of the Transformer encoder-decoder model \cite{vaswani2017attention} and ResNet \cite{he2015deep}, and by leveraging multiple data modalities including text, image, and structures of a UI. We then conducted a Mechanical Turk study to obtain human ratings on summaries generated by different model variants \bw{and heuristic baselines}. \bw {The human evaluation also shows that our full model outperformed all other methods on subjective rating.} Lastly, we discuss potential applications that could benefit from Screen2Words. Altogether, our paper makes the following contributions:
\begin{itemize}
\item We formulate mobile screen summarization, a novel task to automatically generate a descriptive language overview for mobile screens, which expands prior work for screen understanding and language generation by generating language descriptions for the entire mobile screen. The task has important implications for language-based interaction. 

\item We collect, analyze, and open-source\footnote{\url{https://github.com/google-research/google-research/tree/master/screen2words}} the first dataset dedicated for UI screen summarization. It contains 112,085 quality human annotations for 22,417 unique Android screens, collected with a carefully designed labeling process and guideline, which achieve a high inter-labeler agreement for both linguistic coherence and on-screen focus area consistency. 

\item \bw{We develop, train, and evaluate a set of deep models with automatic metrics and human evaluation. The results showed that our full model significantly outperforms all the heuristic baselines and model variants on both automatic metrics and subjective ratings, validating our approach with Screen2Words. The dataset, the models and the empirical results establish a solid benchmark for future research to bridge user interfaces and natural language.}

\end{itemize}




\section{Related Work}
Our work involves both dataset and model development, and builds upon several areas of existing work, including uni- and multi-modal content summarization, mobile screen understanding using deep learning, and mobile UI datasets.

\subsection{Uni- and Multi-Modal Content Summarization}
Automatic content summarization such as text document summarization \cite{maybury1999advances, hovy1999automated} and video captioning \cite{venugopalan2015sequence,chen2017video, dilawari2019asovs} has been widely investigated over the past decades, due to its abundant applications. State-of-the-art summarization techniques typically use deep learning models to encode the underlying representation of the content for generating short, informative text summaries. For example, text summarization \cite{paulus2017deep, yousefi2017text} generates concise summaries for large documents, while image captioning \cite{Showandtell, xu2015show} generates natural language captions to describe input images. Since many real-world data is by nature multi-modal, summarization techniques that use more than one data modality have also been extensively studied \cite{zhu2018msmo,zhu2020multimodal,1221239}. We formulate UI screen summarization as a multi-modal summarization task because it leverages input from multiple data sources, including screenshot images, texts, and structural information of mobile UIs. This task contributes to the spectrum of problems for automatic content summarization and can potentially enhance a class of language-based human-computer interaction problems.

\subsection{Mobile Screen Understanding using Deep Learning} 
There has been an increasing interest in the field \cite{li2021screen2vec, tappability2019, swire2019, bunian2021vins, li-etal-2020-mapping, li2020widget} for using deep models to learn the latent representation of mobile UIs, which we refer to as \textit{screen understanding}. For example, Screen2Vec \cite{li2021screen2vec} uses a self-supervised approach to learn the representation of a mobile UI using the textual content, visual design and layout patterns of the screen, and its app meta-data. Screen understanding has been shown crucial for many downstream tasks. For instance, TapShoe \cite{tappability2019} predicts whether a UI element is tappable by encoding the UI screenshot with CNNs. Similarly, VINS \cite{bunian2021vins} encodes UI designs and wireframes to enable content-based UI retrieval; Swire \cite{swire2019} allows a designer to retrieve UI designs with sketching by encoding both the sketch and UI images. Widget Captioning \cite{li2020widget} and Screen Recognition \cite{zhang2021screen} predict semantically meaningful alt-text labels for GUI components. Screen2Words extends this line of work to predict language summaries for the entire mobile GUI, which requires a model to have a holistic understanding about a screen, and the ability to summarize complex screen contents as a concise language description.

\subsection{Mobile GUI and Interaction Datasets}
Large-scale mobile GUI data repositories are crucial building blocks for data-driven model development. The Rico dataset \cite{Rico, Liu:2018:LDS:3242587.3242650} contains visual, textual, structural, and interactive design properties of 66k unique UI screens from 9.7k Android apps spanning 27 categories in the Google Play Store. ERICA \cite{Deka:2016:EIM:2984511.2984581} provides a collection of user interaction data for mobile UIs captured while using the app. Swire \cite{swire2019} and VINS \cite{bunian2021vins} open-sourced the datasets used for training their UI retrieval models. Another category of work in this area contributes public datasets that connect mobile UIs with natural language for both language grounding and generation \cite{li-etal-2020-mapping, li2020widget}. Based on previous work, our work contributes the first open-sourced, large-scale dataset for mobile UI summarization with high-quality human annotations, detailed analysis and benchmark models.

\section{Dataset Creation}
We start our investigation of methods for automatic screen summarization by creating a dataset, which results in 112,085 human-annotated English summarization for 22,417 unique UI screens. The dataset lays the foundation for data-driven model development for screen summarization. Below we first describe our data collection process and then report an analysis over the collected data. 

\subsection{Mobile UI Corpus}
We started by constructing a mobile UI corpus consisting of screens from an opensource dataset Rico-SCA\footnote{\url{https://github.com/google-research-datasets/seq2act}} \cite{li-etal-2020-mapping}. It contains a subset of screens filtered from the Rico dataset~\cite{Rico} to eliminate screens with missing or inaccurate view hierarchies. In our corpus, each screen comes with a screenshot image of the UI and a view hierarchy JSON file. The view hierarchy is a structural tree representation of the UI where each node, corresponding to a UI element, contains various properties such as the class, visibility-to-user, and bounds of the element. In total, we  labeled 22,417 unique screens from the filtered corpus of Rico-SCA. 

\subsection{Data Annotation}
 \bw{
 We recruited 85 professional labelers to generate the summarization annotation. The labelers were a group of contractors hired for data annotation to assist ML R\&D in our company. All labelers are fluent in English and had previously labeled Android UIs for other tasks. For each screen in the corpus, we presented the screenshot image, task instructions, and the app description to the labeler and collected five annotations from five different labelers.
 Labelers entered their answers in a text field and were able to skip a screen if they found the screen not understandable. On average, each labeler spent around 50 hours creating the annotations. During labeling, a team of quality analysts  sampled and audited around 5\% of the the labels (both screen summaries and SFA) to ensure quality, and incorrectly labeled screens were re-labeled. We trained the labelers with golden examples, pilot data collection, and the following guidelines containing Do's and Don’ts:
}
\\
\\
\indent \textbf{Do's}
    \begin{itemize}
        \item {Summarization should contain 5 to 10 words.}
        \item {Focus on the most important functionalities.}
        \item {Use the texts on the screen to help summarization.}
        \item {Summarize with the structure of “NOUN + CLAUSE”.}
    \end{itemize}

 \textbf{Don'ts}
    \begin{itemize}
        \item {Do not summarize only the images/icons, summarize the whole screen.}
        \item {Do not describe how you feel about the screen, focus on matter of fact.}
        \item {Do not just describe color, shapes and the name of UI elements.}
        \item {Do not mention the name of the app, describe with its category.}
    \end{itemize}

\begin{figure}
     \centering
     \begin{subfigure}[b]{0.46\linewidth}
         \centering
         \includegraphics[width=\linewidth]{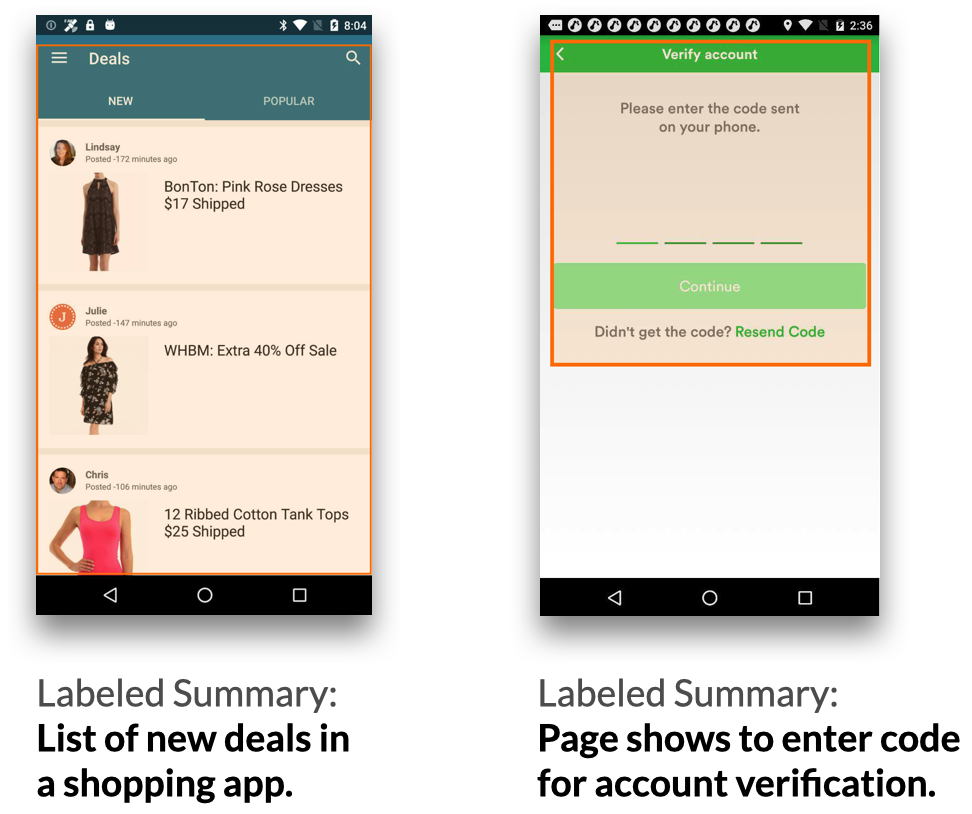}
         \caption{ }
         \label{fig:datasample-a}
     \end{subfigure}
     \hfill
     \begin{subfigure}[b]{0.44\linewidth}
         \centering
         \includegraphics[width=\linewidth]{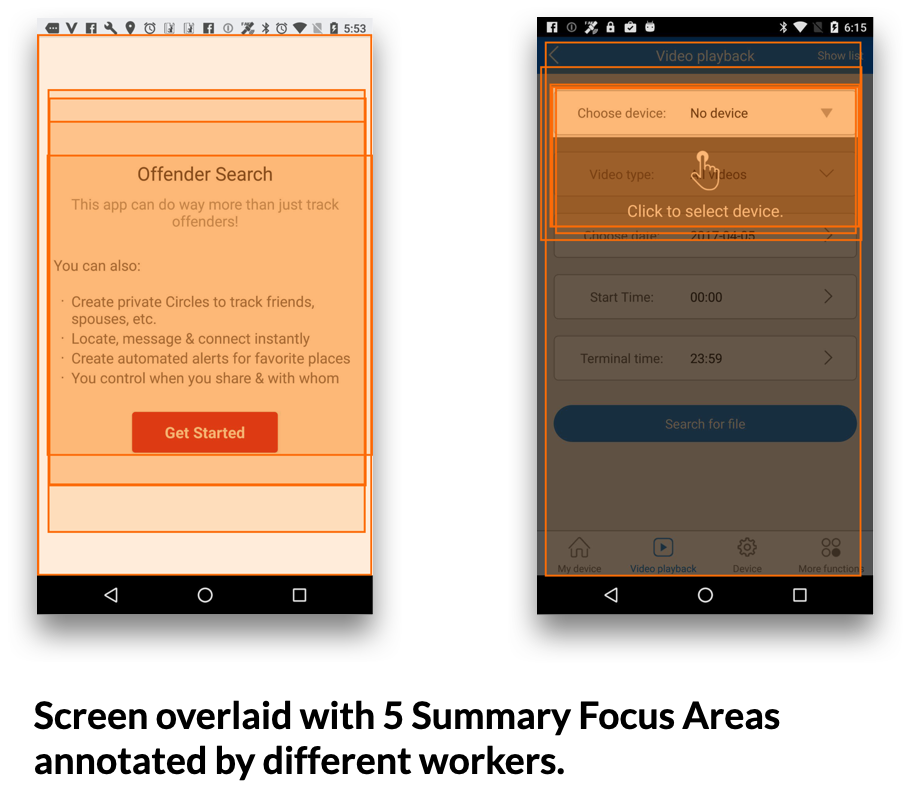}
         \caption{}
         \label{fig:datasample-b}
     \end{subfigure}
     \hfill

        \caption{(a) Data annotations examples. Each screen is annotated with a language summary and a Summary Focus Area (visualized with orange rectangles) indicating where the labeler considered is most influential for their summary. (b) Overlay of 5 labelers' Summary Focus Area. Areas with the deepest colors represent the places where most labelers consider important when summarizing the screen.}
        \label{fig:datasample}
\end{figure}

The guidelines were adapted from those used in the data collection of Microsoft COCO Captions \cite{chen2015microsoft}, a well-known image captioning dataset. Our guidelines are designed to encourage labelers to focus on the functionality of the UI instead of the appearance of screenshots. We iteratively refined the guidelines and the labeling interface based on the results of our pilot studies. To understand the rationale of labelers for composing a screen summary, we also asked them to select a Summary Focus Area (SFA) upon summarizing the screen and entering the phrase. An SFA is an area on the screen covering UI elements that labelers deem are most informative for them to generate the summary.  The SFA annotation focuses on the \textit{summarization importance} for a mobile UI, which is different from the visual importance that previous work~\cite{visualimportance}  focuses on. \bw {We instrumented the SFA to capture potential inconsistency of labeler focus we found in pilot data collection,  e.g., the ads bar versus the main content.}
A labeler can annotate the SFAs simply by marking a bounding box on the UI screen via drag-n-drop. 
Figure \ref{fig:datasample} (a) shows two sample screens with their screen summaries and SFA annotated. 


\subsection{Data Analysis}
\subsubsection{Summary Language Analysis}\label{sec.lang.analysis}
Our dataset contains 22,417 unique screens from 6,269 apps, resulting in \bw{112,085} summary phrases composed by human workers. \bw{We remove a shortlist of stop phrases from collected summary phrases. These stop phrases contain eight variants of \textit{"in the app"}, which often appear in a summary phrase but do not meaningfully contribute to summarization. The average phrase length after stop phrase removal is 6.57 words}. Fig. \ref{fig:word, and phrase analysis} (a) shows the distribution of the summary length we collected. For each screen, to measure how consistent its summaries are across different labelers, we measure the inter-annotator agreement by computing the word-level precision and recall for all the words with two or more occurrences in the collected summaries. The same approach has been previously used in the COCO image captioning dataset \cite{chen2015microsoft}. \bw{The word-level precision and recall are computed using each word’s true positive (TP), false positive (FP), and false negative (FN) counts accumulated across the corpus. For every screen summary, we check whether each of its words appears in the other four summary annotations for the same screen (TP) or not at all (FP). Any word presented only in the other four summaries but not in the one being checked is counted as an FN case. We iterated this calculation through each of the five summary labels for every screen to avoid sampling variance. This process yields corpus-wide accumulative TP/FP/FN counts for each word, which we use for precision and recall calculation.} We focus our analysis on the top 4.5K words in the dataset, which appear more than once and amount to 99.7\% of all the word occurrences in the summaries. Fig. \ref{fig:word, and phrase analysis} (b) shows that the collected summaries have a reasonable word-level agreement for each screen across different human labelers. Specifically, for the 4.5K words, we report the mean precision and recall of every 10 consecutive words in the vocabulary (sorted by word frequency). As a result, the figure contains 450 data points, each representing precision/recall of 10 words. The ranks of the words in the vocabulary are used to color the data points. Lower rank indicates higher word frequencies in the corpus.

\subsubsection{Summary Focus Area Analysis}

The SFAs are approximate indicators of which parts of the UI the labelers focused on when performing summarization, and we use it to understand the perceived importance of UI elements on the screen across labelers for summarization purposes. On average, the SFAs cover 66.1\% of the screen, which is within our expectation as the goal is to summarize the entire screen. We use the IoU score (Intersection over Union) to measure the agreement on the areas labeled by different labelers for the same screen. The average pair-wise IoU score is 74.1\%, showing a decent consensus among labelers on which area on the screen was focused when producing summaries for a screen. \bw {The SFAs were initially introduced to account for potential variance in annotations. As we revised our data collection guidelines and labeling interface through iterations, we found the discrepancy between labelers' focus became less---their SFAs often contain most of the relevant UI elements on the screen. The consistency is desirable for an ML model, and SFA provides a safeguard and accountability for screen summarization consistency.} Fig.~\ref{fig:datasample} (b) shows two sample screens overlaid with the five SFAs annotated by different labelers. Areas with the deepest colors represent the parts where the most labelers consider important when summarizing the screen.

\begin{figure}
     \centering
     \begin{subfigure}[b]{0.45\linewidth}
         \centering
         \includegraphics[width=\linewidth]{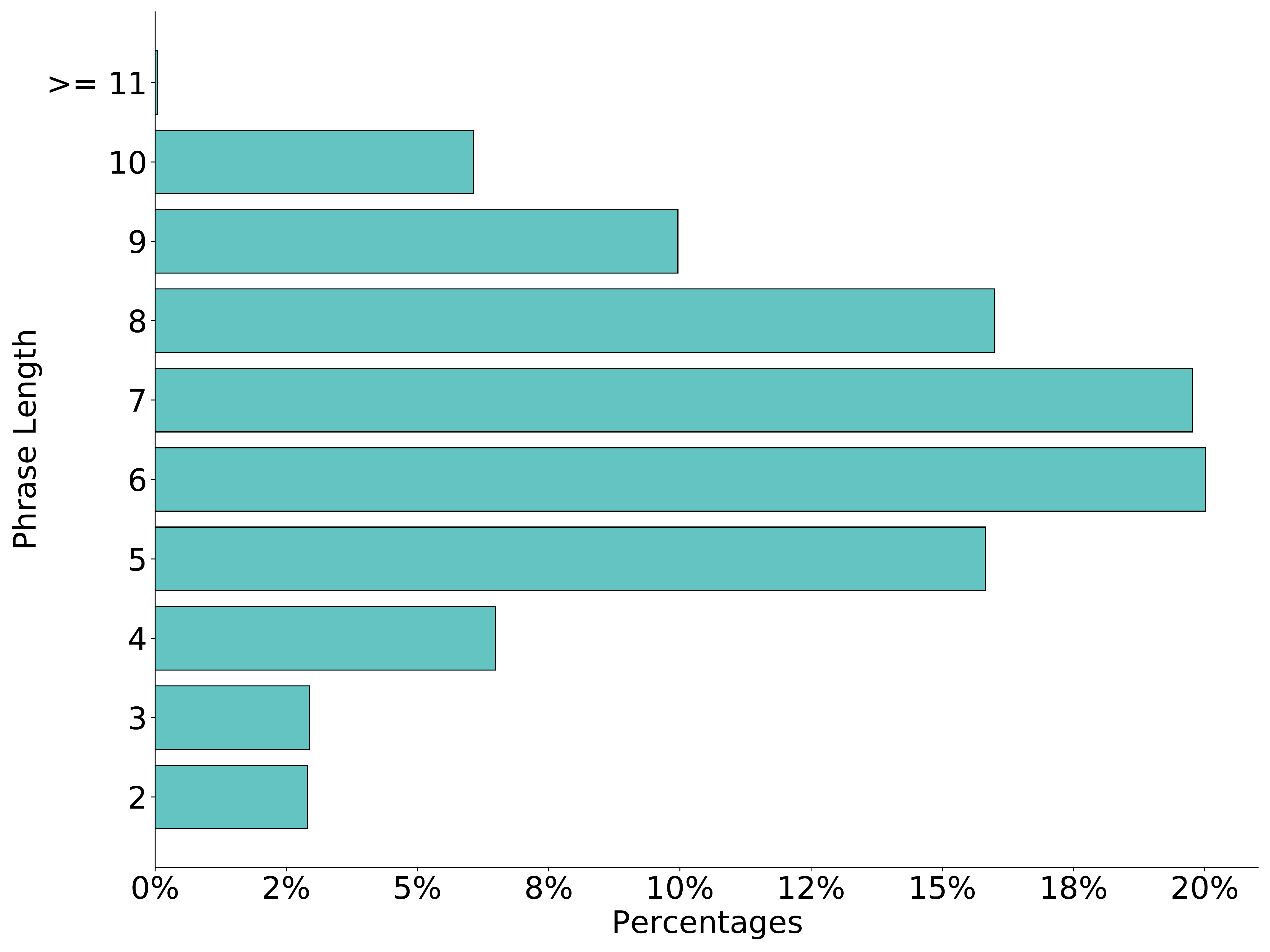}
         \caption{ }
         \label{fig:y equals x}
     \end{subfigure}
     \hfill
     \begin{subfigure}[b]{0.45\linewidth}
         \centering
         \includegraphics[width=\linewidth]{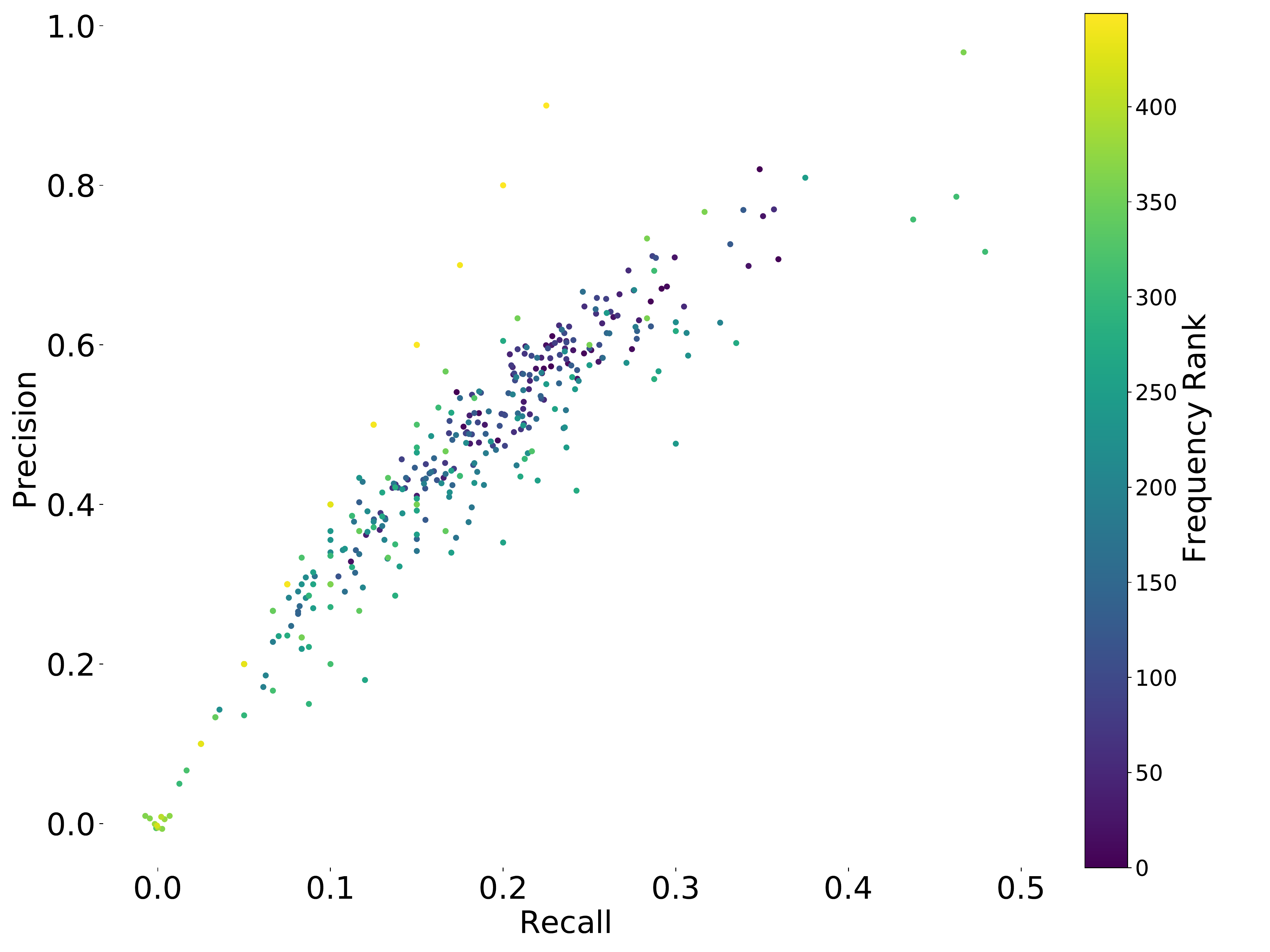}
         \caption{}
         \label{fig:three sin x}
     \end{subfigure}
     \hfill

        \caption{(a) The distribution of summary length of the collected annotations. (b) The distribution of precision and recall for the top 4.5K in the summaries. See section \ref{sec.lang.analysis} for more details.}
        \label{fig:word, and phrase analysis}
\end{figure}


\section{Model Design}
\begin{figure*}[h]
  \hspace*{0.05cm}   
  \centering
  \includegraphics[width=0.97\linewidth]{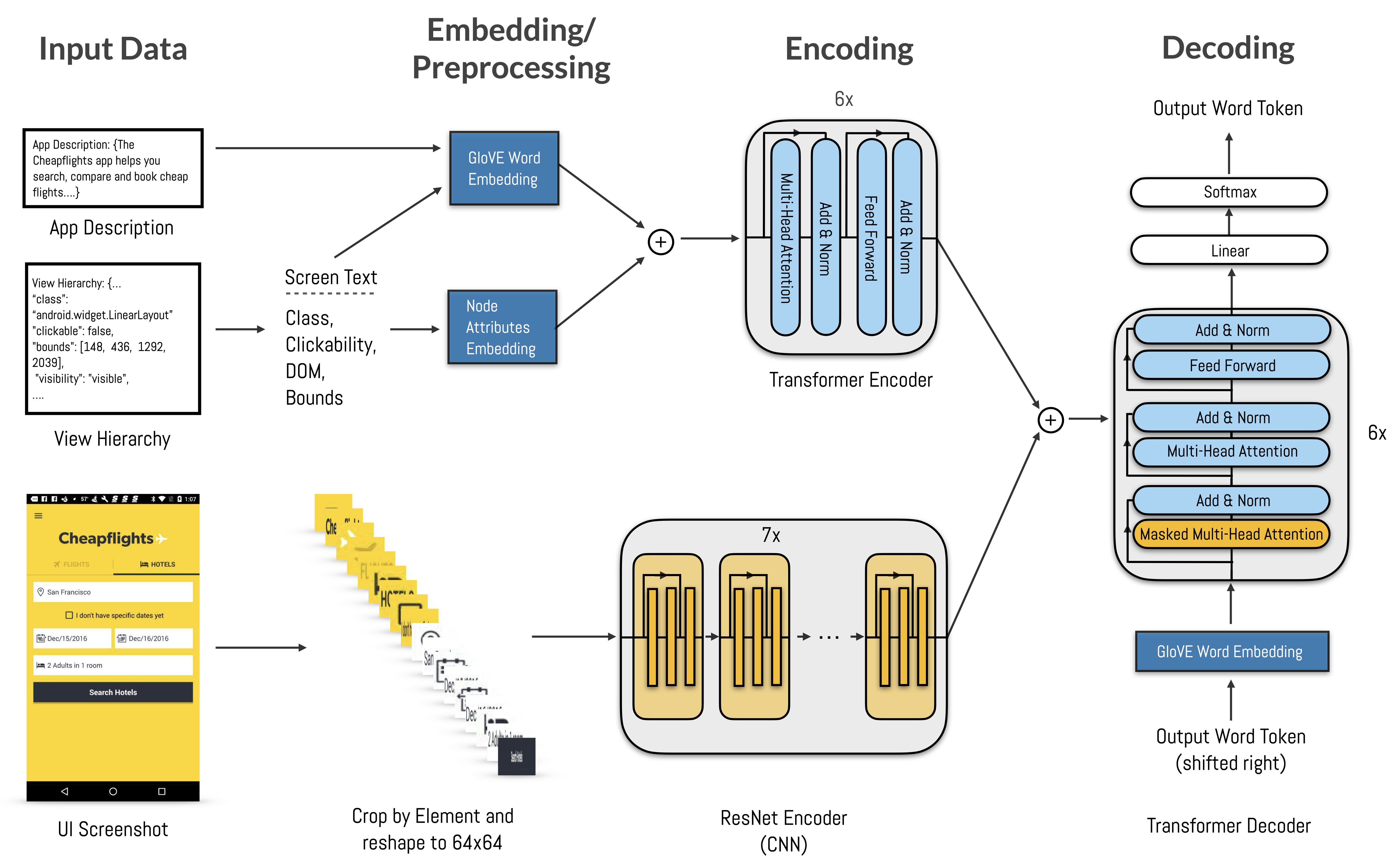}
  \caption{The Screen2Words deep model is designed based on the Transformer Encoder-Decoder architecture~\cite{vaswani2017attention}. It leverages three input modalities, including the UI screenshot, the view hierarchy structure, and the screen text and app description. Each modality is encoded and then fused to participate in the decoding for summary generation.}
  \Description{model architecture}
  \label{fig:architecture}
\end{figure*}
We utilize deep learning models to understand the challenges and feasibility of the proposed screen summarization task. We designed our models based on the encoder-decoder architecture, a commonly used architecture in image captioning and machine translation. In an encoder-decoder architecture, the model first encodes input data into hidden representations---referred to as encoding, and then decodes outputs based on the encoded information.

Figure~\ref{fig:architecture} shows the architecture of our screen summarization model. Similar to models of Widget Captioning~\cite{li2020widget}, we use a dual encoder for encoding multi-modal information of a screen. Our encoder consists of 1) a Transformer encoder to encode the UI structure of a screen along with the text description of the app that the screen belongs to, and 2) a ResNet to encode the image pixels of each element on the screen. The outputs of the two encoders are then combined via a late fusion, i.e., the concatenation of the outputs of the two encoders forms the encoding of each element. The resulted encodings of all the UI elements are the multi-modal semantic understanding of the screen, which is then fed to the Transformer decoder for generating the phrase for screen summarization. We next elaborate on each individual component of our model. Although many aspects of our model designs follow best practice, we would like to describe them here for completeness and reproducibility. Meanwhile, we highlight unique aspects in our design, such as incorporating app description as an additional "element" to participate in Transformer self-attention during encoding.

\subsection{Encoding a Mobile UI Screen}
To form a holistic understanding of a mobile UI screen, we encode both the structural-textual information in its view hierarchy and app description, and the raw pixels of its screenshot, using a dual encoder. The outputs of each encoder are then concatenated to form the multimodal encoding of each UI element on the screen.
 
\subsubsection{Encoding Structural and Textual Information}
We use a Transformer model \cite{vaswani2017attention} to encode both the structural and textual information of a mobile UI. The core idea behind the Transformer model is self-attention, which the model learns to represent each screen element by leveraging the information from all the elements coexisting on the screen via neural attention mechanisms. As a result, the approach allows us to acquire a contextual representation of each UI element on the screen.  A Transformer encoder requires both content embedding and positional encoding as input, and we follow the approach that has been shown effective for embedding the mobile~\cite{li2020widget, li-etal-2020-mapping}. \bw{We perform breadth-first traversal to flatten the view hierarchy tree and obtain the linear order of input elements required by the Transformer model. We compute the positional embedding using the elements' spatial positions on the screen and the tree (DOM) positions in the view hierarchy. Each element of the flattened tree carries both spatial and structural positional information.}

\textit{Element Embeddings}: We first need to represent each individual element on the UI as an embedding vector. An element comes with a rich set of properties. From view hierarchy, the {\it class}, the {\it clickability} and the {\it bounds} of an element respectively represent the type of the element, whether it is clickable, and its 2D positional information on the screen. The tree positions of the element, captured as the pre-order and post-order traversal position and its depth in the view hierarchy, reflect the structural position of the element on the UI. Both the spatial and tree positions of the element constitute the positional information of the element. Last but not the least, an element might come with text content, which is a good source of semantic information about the element. We embed each of these sources of information separately for an element. For the text content, we use pre-trained GloVe word embedding \cite{pennington-etal-2014-glove} to represent each word and then perform sum pooling over all the words of the element to acquire a fixed-size vector for the element. For all other types of attributes, we treat them as categorical variables and embed them separately. The final embedding of an element is acquired via concatenation. 

\textit{App Description Embedding}:  As described in our data collection section, we allow the human labelers to read the app description while performing screen summarization, because the app description provides the background information about the app that a screen belongs to. To train a computational model to achieve human intelligence in screen summarization, it is important to give the model the same access to this information. As such, we include the app description from Google Play Store as another source of input. Similar to representing the text content in an element, we first embed each word in an app description using pre-trained GloVe word embedding, and then simply treat the app description as a "bag of words" and acquire the final embedding of the app description using sum pooling over all the words. For screens with missing app descriptions, we assign an \bw{all-zero embedding} embedding for an empty description. Note that more complex methods, such as LSTM, can be easily plugged into our model here for embedding app descriptions, although it is not the focus of this paper. The outcome of this process is to obtain a fixed-size embedding vector for the app description.

\textit{Transformer Encoding}: With each element and the app description represented as an embedding vector, we linearly project ($P$) each embedding vector to a target dimension that is required by the Transformer encoder, as shown in Figure~\ref{fig:architecture}. Note that we treat the app description as a special "element" to participate in the Transformer encoding process. This design naturally allows each UI element to reference the app description embedding when forming the element's contextual representation, due to the self-attention mechanism of the Transformer encoder model. Specifically, we append the app description embedding to element embeddings, over the element dimension, which yields an embedding tensor, $E$, in the shape of \texttt{[num\_of\_elements + 1, embedding\_size]}. To enable the Transformer encoder to differentiate the embedding vectors of a real UI element from that of the app description, we tag each embedding to indicate whether it is an element embedding or app description embedding, simply by appending a one-hot vector to the embedding before the linear projection $P$. The embedding tensor, $E$, was then fed to the Transformer encoder to produce the structural-textual encoding of the UI.

\subsubsection{Encoding the UI Screenshot}
To represent the visual aspect of a screen, we encode the image of each element on the screen. To do so, we crop the image of each element from the UI screenshot, and then re-scale the cropped image to a fixed-size tensor in the shape of \texttt{64x64x1}, where \texttt{64x64} are the spatial dimensions and \texttt{1} is the single channel for grayscale. \bw{We resize the images to ensure each element's image encoding has a fixed shape expected by the downstream computation. While resizing loses the aspect ratio, it makes our model more efficient in computation compared to padding element images to a larger target size. We use grayscale images because the semantics of most UI elements are color-invariant.} Because the view hierarchy that specifies each element on the screen is given, it is sensible to directly use the cropped element image instead of asking the model to encode the entire screen. This design frees the image encoder from having to learn to extract UI elements from the screenshot pixels---an object detection problem---which by itself is a nontrivial task. We used ResNet, a multi-layer Convolutional Neural Net (CNN), to encode the pixel information of each UI element. The building block of our ResNet is a residual block consisting of three convolutional 2D layers with a residual connection--—the input of the $1$st layer is added to the input of the $3$rd layer. The last layer of each block uses a stride of 2 that halves both the vertical and horizontal spatial dimensions. The output of the multi-layer CNN represents the visual encoding of the elements on the screen. We then concatenate the structural-textual encoding, from the previous section, and the visual encoding to form the \textit{final encoding} of each UI element. Note that because the structural-textual encoding has an extra element from the app description, we add a padding image encoding as its corresponding visual encoding. \bw{We used the padding encoding since the image encoding of each element is already made available to the model. An alternative approach is to use the image encoding of the entire screen.}

\subsection{Decoding Screen Summaries}
The final encodings acquired from the previous section encompass both the structural-textual and visual information about each element on the screen. Such a representation is contextual as each encoding is computed by attending to other elements on the screen. Based on these encodings, we use a Transformer decoder \cite{vaswani2017attention} model to generate a varying-length natural language summary for the screen. Similar to the encoder, the Transformer decoder uses self-attention to attend to the context of the generated word tokens during summarization. To avoid information leaks from future tokens that have not been generated during supervised training, it uses masked self-attention to allow its multi-head attention to only attend to previous token representations. In addition, the Transformer decoder accesses the encoder's output, i.e., the encodings, for each step of the decoding process. This is achieved with the encoder-decoder attention layer. Internally, it adds the weighted sum of screen encodings to the attention output of each decoding step, before feeding into a position-wise, multi-layer perceptron (FFN). Unlike prior work focusing on captioning each individual element on the screen \cite{li2020widget}, our Transformer decoder attends to every element on the screen to decode a holistic screen summary. The probability distribution of each token of the summarization is finally computed using the softmax over the Transformer decoder output. The entire model, including both the encoders and decoder, is trained end-to-end, by minimizing $\mathcal{L}$, the average cross-entropy loss for decoding each token of each screen's summary: 
\begin{displaymath}
  \mathcal{L} =\frac{1}{N}\sum_{i=1}^{N}\frac{1}{M}\sum_{j=1}^{M}\mbox{Cross\_Entropy}(y'_{i,j} , y_{i,j}),
\end{displaymath}
where $M$ is the number of word tokens to decode and $N$ is the number of screens in the mini batch, $y'_{i,j}$ is the $j$th token in the groundtruth summary and $y_{i,j}$ is the corresponding prediction. Training is conducted in a teacher-forcing manner where the groundtruth summary words are fed into the decoder. During prediction time, the model decodes autoregressively and beam search is used to generate the top summarization candidates.

\begin{table}
  \caption{Experimental Dataset statistics}
  \label{tab:dataset_stats}
  \begin{tabular}{lccc}
    \toprule
    Dataset & \#Apps & \#Screens & \#Summaries\\
    \midrule
    Training & 4,390 & 15,743 & 78,715\\ 
    Validation & 625 & 2,364 & 11,820\\
    Test & 1,254 & 4310& 21,550\\
    \midrule 
    Total & 6,269 & 22,417& 112,085\\
  \bottomrule
    \end{tabular}
\end{table}

\begin{table*}
  \caption{Model Performance on Automatic Metrics}
  \label{tab:performance}
  \begin{tabular}{lcccccccc}
    \toprule
        Model & BLEU-1 & BLEU-2 & BLEU-3 & BLEU-4 & CIDEr & ROUGE-L &  METOER  \\
    \midrule
        Template (TF-IDF) & 47.9 & 29.3 & 20.7 & 16.5 & 31.8 & 36.5 & 21.1  \\
        Template (Pixel) & 42.8 & 25.3 & 17.9 & 14.6 & 14.1 & 33.9 & 18.6  \\
        Template (Pixel-DL) & 43.6 & 26.0 & 18.3 & 14.8 & 16.1 & 34.2 & 19.1  \\
        Template (TF-IDF+Pixel+AppDesc) & 45.7 & 27.3 & 19.2 & 15.5 & 23.7 & 35.5 & 20.1  \\
        Template (TF-IDF+Pixel) & 49.0 & 30.0 & 21.1 & 16.8 & 33.2 & 37.6 & 21.5  \\
        Pixel Only& 56.8 & 37.0 & 25.3 & 19.9 & 31.4 & 42.6 & 25.6 \\ 
        Layout Only&  58.7 & 39.6 & 27.4 & 21.7 & 35.2 & 44.3 & 26.0 \\
        Pixel+Layout & 62.1 & 40.6 & 28.4 & 22.1 & 35.4 & 45.5 & 26.1 \\ 
        Pixel+Layout+ScreenText & 63.6 & 43.5 & 30.9 & 23.9 & 55.5 & 47.4 & 29.0 \\ 
        Pixel+Layout+ScreenText+AppDesc & \textbf{65.5} & \textbf{45.8} & \textbf{32.4}& \textbf{25.1}& \textbf{61.3} & \textbf{48.6} & \textbf{29.5}  \\ 
  \bottomrule
    \end{tabular}
\end{table*}

\section{Experiments}
We conducted screen summarization experiments to investigate the effectiveness of our proposed Screen2Words approach based on multi-modal deep learning. The goal of the experiments is to validate whether 1) deep models perform better than heuristics methods and 2) whether incorporating multiple data modalities would lead to better summarization results. We first discuss the experimental setup and training configurations. We then report the performance of our models against several commonly-used metrics and an analysis of the model behavior.

\subsection{Experimental Datasets}
We split our dataset into training, validation and test set for model development and evaluation, as shown in Table~\ref{tab:dataset_stats}. To avoid information leaks because screens in the same app might share similar styles and semantics, we split the data app-wise so that all the screens from the same app will not be shared across different splits. Consequently, all the apps and screens in the test dataset are unseen during training, which allows us to examine how each model configuration generalizes to unseen conditions at the test. Our vocabulary includes 10,000 most frequent words and the rest of the words encountered in the training dataset is assigned a unique unknown token \texttt{<UNK>}. During validation and testing, any \texttt{<UNK>} in the decoded phrase is removed before evaluation. Since in our dataset, each screen has five summarization label phrases (from five different labelers), one of its labels is randomly sampled as the training target each time during stochastic training. During the testing phase, all the five screen summaries are used to form the reference set for automatic evaluation metrics such as BLEU and CIDEr. 

\subsection{Model Configurations and Training Details}
We tuned our model architectures based on the training and validation datasets. We initialize the word embeddings with pre-trained 400K-vocab 300-dimensional GLOVE embeddings \cite{pennington-etal-2014-glove}, which are then projected onto a 128-dimensional vector space. The embedding weights were shared by both the screen encoder and the decoder. Both the Transformer encoder and decoder use 6 Transformer layers with a hidden size of 128 and 8-head attention. A 7-layer ResNet was used for encoding the pixels of elements on the screen. The ResNet in total involves 21 convolutional layers, and the output of the final layer is flattened into a 256-sized vector. Batch normalization was used for each convolutional layer. The final encoding of each element's image is a 128-dimensional vector that is concatenated with the Transformer encoder's output for decoding. Each word during summarization is decoded sequentially, and we use beam search with a beam size of 5 to generate top-5 summarization predictions during testing. We implemented the models with TensorFlow, and trained the models with 8 Tesla V100 GPU cores using a batch size of 128 screens, and the Adam optimizer \cite{kingma2017adam}. All the models we experiment with were converged in less than two days.

\subsection{Model Variants and Baseline}
To investigate the effectiveness of fusing multi-modal data representation for screen summarization, we compared different variants of our model:
\begin{itemize}
    \item \textit{PixelOnly}. Building upon the existing image captioning approach \cite{mao2014deep, Showandtell, xu2015show}, this variant leverages only the visual information encoded by the multi-layer CNN from the UI element images to generate summaries.
    \bw{\item \textit{LayoutOnly}. This variant uses the structural representation encoded with the Transformer encoder based on the UI properties of the elements extracted from the view hierarchy.
    \item \textit{Pixel+Layout}. This variant uses both the image encoding and structural information.}
    \item \textit{Pixel+Layout+Text}. Based on Pixel+Layout, this model additionally uses the textual information encoded from the screen texts.
    \item \textit{Pixel+Layout+Text+AppDesc}. This is our full model that uses all the above information plus the app descriptions.
\end{itemize}

 Since there are no existing baseline techniques dedicated to automatic screen summarization, we create several template-based baselines for comparison in our experiment. The baselines predict the summary of an unseen screen---the query screen---by retrieving the summary of the screen, from the training dataset, that is most similar to the query screen---the nearest neighbor approach. In these baseline methods, we featurize each screen as either a TF-IDF score vector, which is well-known in the literature of text analysis, or a vector of pixel values, or a combination of the two.
 
 
 The length of the TF-IDF vector is the vocabulary size. Each value in the vector corresponds to the TF-IDF score of a word in the vocabulary. Based on the original TF-IDF terminology, we treat each screen as a \textit{document} and each word in the screen as a \textit{term}. The Term-Frequency (TF) represents how often a word token appears in the screen, whereas Inverse Document Frequency (IDF) tells how rare the word is across screens in the training corpus. Together, a TF-IDF score measures how unique a word is to the screen. We implement the TF-IDF baseline using sklearn \cite{sklearn_api}. For the pixel-value vector, we convert the screen to grayscale and resize the screen to a dimension of $100 \times 100$ to match the TF-IDF vector size.  \bw{Additionally, we include \textit{Pixel-DL}, another pixel-based baseline that uses image encoding learned by a CNN-based autoencoder\footnote{The Pixel-DL baseline uses a CNN-based autoencoder trained to recover the screenshot image with mean square loss. The model consists of 3 convolutional layers (kernel sizes: 128, 64, 32) and 3 transposed convolutional layers (kernel sizes: 32, 64, 128), all with a filter size of 3x3 and strides of 2. The latent state of a screenshot is a vector of size 100, which is used to retrieve the most similar screenshot and its human-labeled summary.} to represent the pixel-value vector.}  
 With each screen represented as a TF-IDF vector and/or a pixel-value vector, we can then compute the cosine similarity scores $\in [0,1]$ between a pair of screens, and find the most similar screens using either scores or the sum of the two scores. We investigate baseline variants using either or both vectors to match a query screen to screens in the training set, and report their performance in Table \ref{tab:performance}. \bw{The baselines allow us to understand the challenge of using a traditional nearest-neighbor or template-based retrieval method and the benefit of using a deep generative model for automatic screen summarization.} In the following reports, we prefix each baseline method with \textit{Template} because they are all based on template-based approaches.
 
 \begin{figure*}[h]
  \centering
  \includegraphics[width=0.95\linewidth]{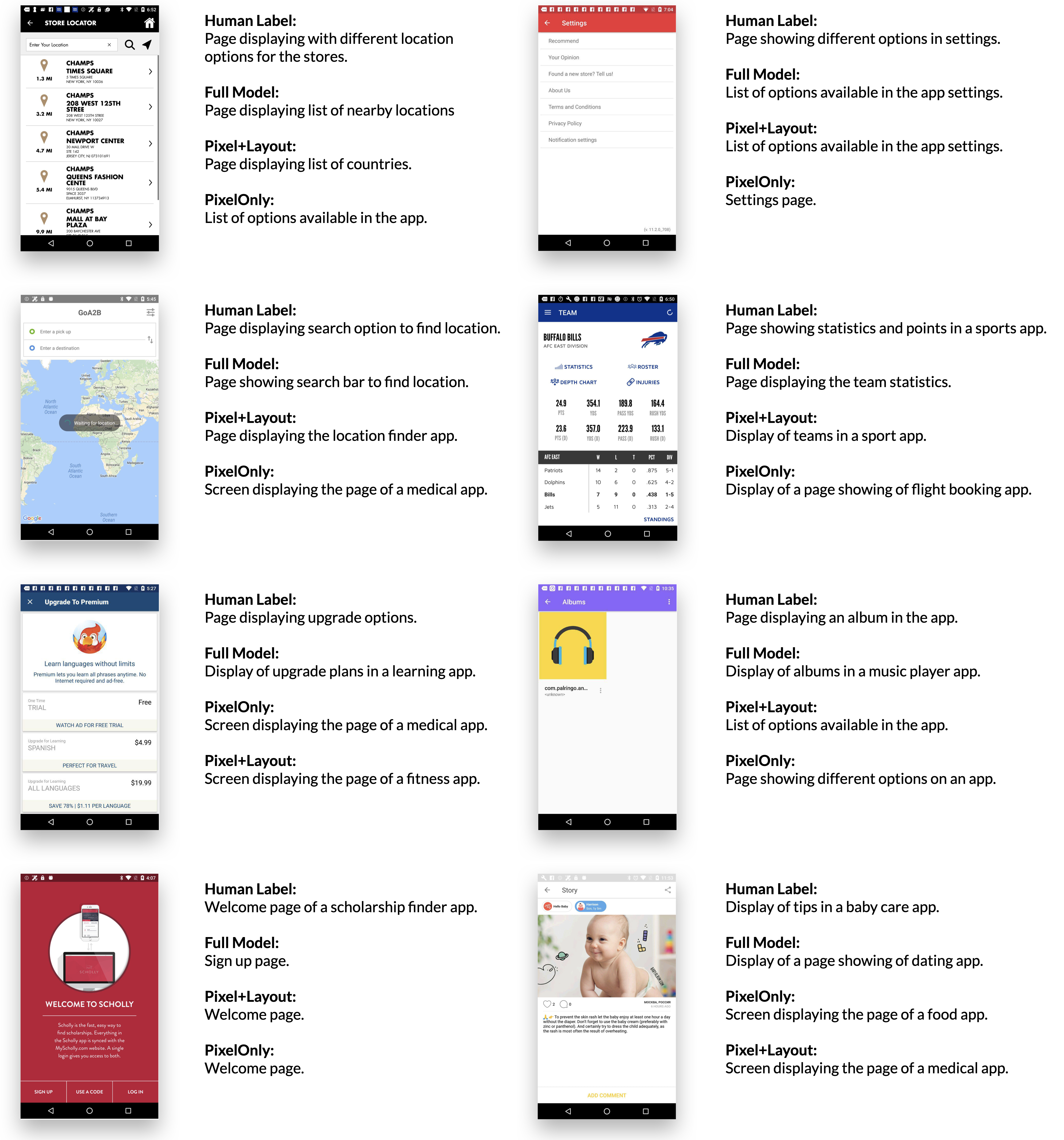}
  \caption{Summarization examples generated by our models. All the examples are sampled from the test set.}
  \Description{Model Results Examples.}
  \label{fig:model_results}
\end{figure*}

 
\subsection{Experimental Results}
In this section, we report our model performance based on metrics commonly used in machine translation and image captioning tasks: BLEU \cite{papineni2002bleu}, CIDEr \cite{vedantam2015cider}, ROUGE-L \cite{lin2004rouge}, and METOER \cite{denkowski2014meteor} (see Table~\ref{tab:performance}). A higher number means better model performance for these metrics---the closer distances between the predicted and the ground truth phrases. For all model variants, the top-1 prediction by beam search is used for calculating metrics; for the template-based baselines, we sample one summarization from the most similar screen for its prediction. All the scores were calculated on the test dataset, which was not seen by the model during the training phase.

The results show that all the \bw{generative models outperformed all the template-based retrieval baselines} across these metrics by a large margin, which justifies our proposed \bw{modeling} approach. Among the baselines, using both TF-IDF and pixel-value vectors achieved the best performance, demonstrating the usefulness of multimodal information. Adding app description for computing the TF-IDF vector (Template (TF-IDF+Pixel+AppDesc) in Table \ref{tab:performance}) did not help the performance, possibly due to that all the screens from the same app would have the same app description, making them indistinguishable from each other if the screen text is sparse. \bw{The Template (Pixel-DL) baseline, which uses DL-based image encoding, offers only modest improvements over using raw pixels.}

\bw{On the other hand,} the Pixel-Only model, which only uses the image encoding of an element, performs significantly better than all the baselines. 
\bw{The Layout-Only model, which incorporates the mobile UI structural representation, achieves slightly better performances than the Pixel-Only model. Combining the two inputs, Pixel+Layout offers further improvements across the metrics.} Similarly, Pixel+Layout+ScreenText, which additionally leverages the textual information from screen, achieves even better results. Our full model, which leverages app description, achieves the best accuracy among all the models. These results show that the multi-modal data of mobile UI complement each other, and their combination leads to better summarization results.

Figure~\ref{fig:model_results} shows example summarization results by different model variants on the same test set screens. All of our models are able to compose coherent, understandable summaries, with our full model having the most substantial capability to capture the underlying semantics of a screen and generate the most accurate summaries. While the Pixel-Only model can sometimes generate relevant descriptions (row 1), it often fails to capture the meaning of a more complex screen. Examples in row 2 demonstrate that additional layout information does help the Pixel+Layout model boost the summarization performance. Our full model results reveal that textual information is helpful to provide more contexts and details when summarizing a screen, such as app categories (row 3). To showcase when the model may fail, we also include examples where all models could only generate very generic descriptions (row 4, left), and where all of them were unable to generate a relevant summary (row 4, right).

\section{Human Evaluation}
To further understand the quality of screen summarization, we conducted a Mechanical Turk study to ask human to assess the quality of the generated screen summary, and validate how automatic metrics correlate with human judgment.

\subsection{Study Setup}
We compare the model variants \textit{Pixel Only}, \textit{Pixel+Layout}, and the full model \textit{Pixel+Layout+ScreenText+AppDesc}, \bw{as well as two template baselines \textit{Template (Pixel)} and \textit{Template (TF-IDF+Pixel)}} on the same set of 1000 screens that we randomly sampled from the test set. For each screen, we recruited three raters to assess the quality of summarization. In total, there were 1041 unique human workers participated in the rating, and none of them was involved in the dataset labeling.  During rating, human raters were presented with a screen summary generated by one of these models along the corresponding UI screenshot. Human raters are not aware of which model generates the screen summary being evaluated. We asked the raters to consider three aspects:  

\begin{itemize}
    \item \textit{Screen Type}: If the summary mentions the type of screen,  e.g., sign-in or sign-up screen, settings menu, search results, how accurate or relevant is it?
    \item \textit{UI Elements}: If the summary mentions UI elements, e.g., pop-up, list, options, search bar, video, how accurate or relevant is it?
    \item \textit{App Type}: If the summary mentions the type of the app, e.g., social app, news app, learning app, how accurate or relevant is it? 
\end{itemize}

For the rating scales, we adopted the rating system previously used in evaluating image captioning quality with human judgments~\cite{Showandtell}, but added the fifth point, which indicates the summary not only describes the screen without errors but also provides useful details of the screen, instead of a generic description. The exact rating criteria provided to the labelers are as follows:
\begin{itemize}
    \item describes the screen without any errors and provides sufficient details.
    \item describes the screen without any errors.
    \item describes the screen with minor errors.
    \item describes the screen with a somewhat related description.
    \item describes the screen with an unrelated description.
\end{itemize}

\subsection{Study Results}
For each screen, we averaged the ratings from three different raters to obtain a score for the summarization quality. \bw{As shown in Table 3, the mean score of each deep model variant from human evaluation correlates with the results of automatic evaluation metrics well. The full model is rated the best, which is followed by the \textit{Pixel+Layout}, and then the \textit{Pixel-Only}. The \textit{Template (Pixel)} baseline has the lowest average rating. Interestingly, the strongest baseline on automatic metrics, \textit{Template (TF-IDF+Pixel),} achieves a higher human rating than the two deep model variants that do not use textual information, even though these deep models outperformed the baseline method on automatic metrics as discussed previously. We speculate this discrepancy between subjective ratings and automatic metric scores might be due to the following. The ScreenText feature may contain keywords that play a decisive role in determining human-perceived relevance of a screen and its summary, which is utilized by TF-IDF+Pixel but deprived from these two deep model variants. The disadvantage does not show in automatic metrics as the metrics treats each word equally instead of weighting them according to their salience to human perception about summarization relevance.

Nonetheless, the results of human evaluation, combined with the automatic metric evaluation, clearly show that our full model outperforms all the template baselines and deep model variants. A non-parametric Mann-Whitney U test shows that the differences between the mean ratings of our full model and all the other settings are significant ($p < 0.0001$). Moreover, 38.6\% of the ratings on the full model received the highest score, i.e., 5 points, and the percentage drops to 32.2\% and 29.5\% for the \textit{Template (TF-IDF+Pixel)} and the \textit{Pixel+Layout}, respectively. These indicate that compared to baselines, our full model is more capable of generating summaries that are not only without any errors but also contain sufficient details.}

\begin{table}
  \caption{Mean Summarization Scores Rated by Mechanical Turk Workers.}
  \label{tab:freq}
  \begin{tabular}{lccc}
    \toprule
    Model & Mean & STD \\
    \midrule
    Template (Pixel) & 2.583 & 1.075\\
    Template (TF-IDF+Pixel) & 3.240 & 1.081\\
    Pixel-Only & 2.886 &  1.180 \\
    Pixel+Layout & 3.166 & 1.060 \\
    Pixel+Layout+ScreenText+AppDesc & \textbf{3.436*} & 1.119 \\
  \bottomrule
    \end{tabular}
\end{table}

\section{Potential Applications}
We have introduced Screen2Words, an approach for automatic screen summarization based on multimodal UI information, and demonstrated the capability of our \bw{proposed model} for generating quality screen summaries. We now demonstrate the potential use cases of Screen2Words by proposing three mock-up applications: 1) Language-Based User Interface Retrieval, 2) Enhancing Screen Readers, 3) Screen Indexing for Conversational Mobile Applications.

\subsection{Language-Based User Interface Retrieval}
\begin{figure}[h]
  \includegraphics[width=\linewidth]{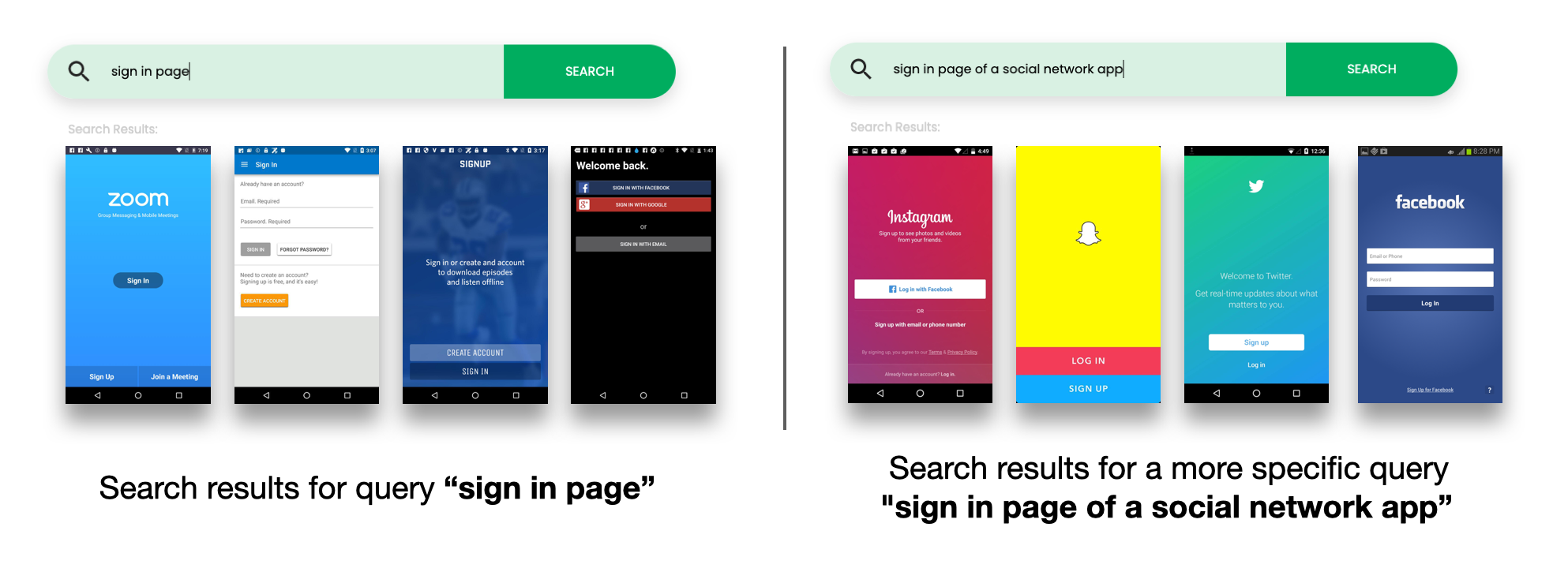}
  \caption{Mock-up interface of language-based UI retrieval system based on Scree2Words. Left: the designer searches for UI design example using the query "sign in page". Right: the designer refines the query to search for sign-in pages of a social network app.}
  \Description{UI retrieval examples.}
  \label{fig:ui-retrieval}
\end{figure}

Screen2Words can empower design search by enabling language-based mobile UI retrieval with which a designer can retrieve design examples using natural language queries. Unlike keyword-based retrieval, which retrieves UIs by directly matching the words in the query with those on the screen texts, Screen2Words models can capture UI semantics beyond text content that appears on a screen, evident by our summary results and accuracy achieved. Figure \ref{fig:ui-retrieval} shows a mock-up interface demonstrating a language-based UI retrieval system. With the UI semantics captured by Screen2Words, designers can not only search with UI keywords like "sign up page" (Figure \ref{fig:ui-retrieval}, left), but also use queries that contain more specific semantic details, such as "sign up page of a social network application" (Figure \ref{fig:ui-retrieval}, right). We use the mock-up to map out the envisioned user experiences of UI retrieval based on Screen2Words. A fully functioning system could be built by training the Screen2Words models with a triplet loss based on our dataset.

\subsection{Enhancing Screen Readers}
Screen readers, e.g., VoiceOver and TalkBack, render text and image content on a mobile screen into speech based on the metadata describing the UI. They are essential accessibility features for a visually impaired person yet often suffer from the missing metadata \cite{10.1145/3234695.3236364, 10.1145/3132525.3132547, 10.1145/3242587.3242616}, which have motivated recent work in predicting missing metadata using machine learning \cite{li2020widget, zhang2021screen}. Screen2Words can contribute to this effort by predicting screen summaries to provide an overview description for screen reader users. Unlike sighted users who can quickly understand the purpose of an unknown screen with the rich visual interface, visually impaired users have to scan through the elements on a screen, with a screen reader, before they form a mental model for a new UI. Therefore, studies \cite{ChallengesScreenReader_Rodrigues, Kuber2012DETERMININGTA, Borodin2010MoreTM} have called out for the need of screen overviews for visually-impaired users so that they can quickly decide whether they should spend time on the current page. 

\begin{figure}[h]
  \centering
  \includegraphics[width=1\linewidth]{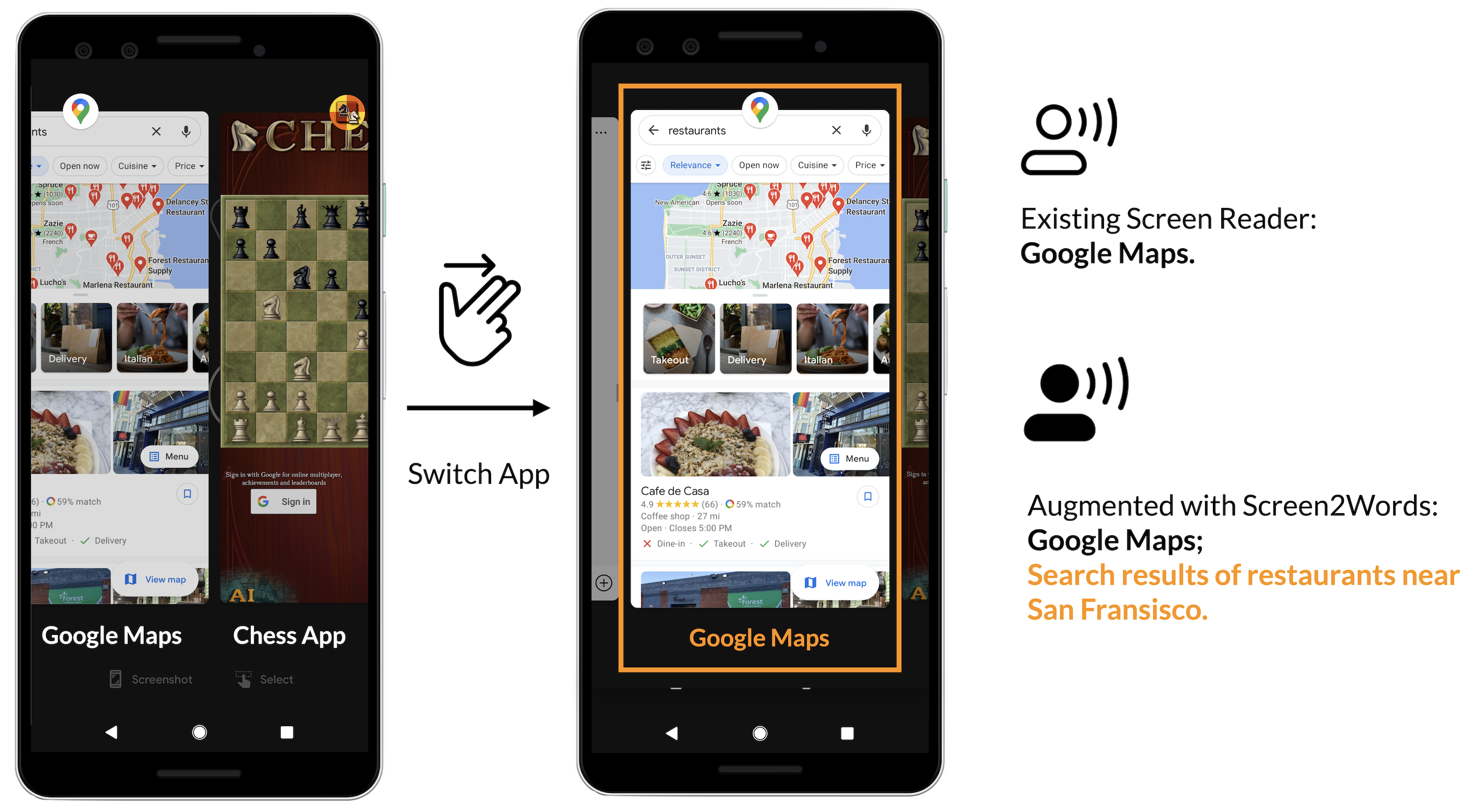}
  \caption{Illustration of how Screen2Words can enhance existing screen readers when using apps switcher.}
  \Description{Mock up 2}
  \label{fig:mock-up-2}
\end{figure}

For a more concrete scenario, when a user switches apps with the app switcher (e.g., Android Overview or iOS App Switcher), the OS typically caches the last viewed screen for each app so that the user would not need to always start from the home screen. However, since visually impaired users cannot see the visuals and the screen readers only read the name of each app, there is no way they could know which specific screen is cached for each app. As shown in Figure \ref{fig:mock-up-2}, Screen2Words can provide a screen summary to help them locate where they are in an app immediately after switching to the app. 

\subsection{Screen Indexing for Conversational Mobile Applications}
Screen summarization can be used to equip each screen with additional language metadata so that they can be indexed on the phone. By combining with other app metadata, the user could quickly launch a desired screen by saying \textit{"Setting page of Gmail"} or \textit{"Ordering page of the Starbucks app"} without manual navigation. Moreover, Screen2Words can be integrated into a multi-modal conversational agent that cooperates with users to accomplish mobile tasks \cite{Toby2019, Toby2018, mmr_li2020}. For example, SUGILITE \cite{sugilite2017} learns to carry out novel tasks by user demonstration. With the screen indexing powered by Screen2Words, speech interaction can replace part of the manual demonstration. For instance, if the user wants to demonstrate how to order an Americano in the Starbucks app, they can say \textit{"Open the ordering page of the Starbucks app."} Once the ordering page is opened, the user can continue to demonstrate the remaining steps manually. The hybrid approach is valuable especially when manually navigating into the desired screen, for completing the task, takes a longer time.

\begin{figure}[h]
  \centering
  \includegraphics[width=1\linewidth]{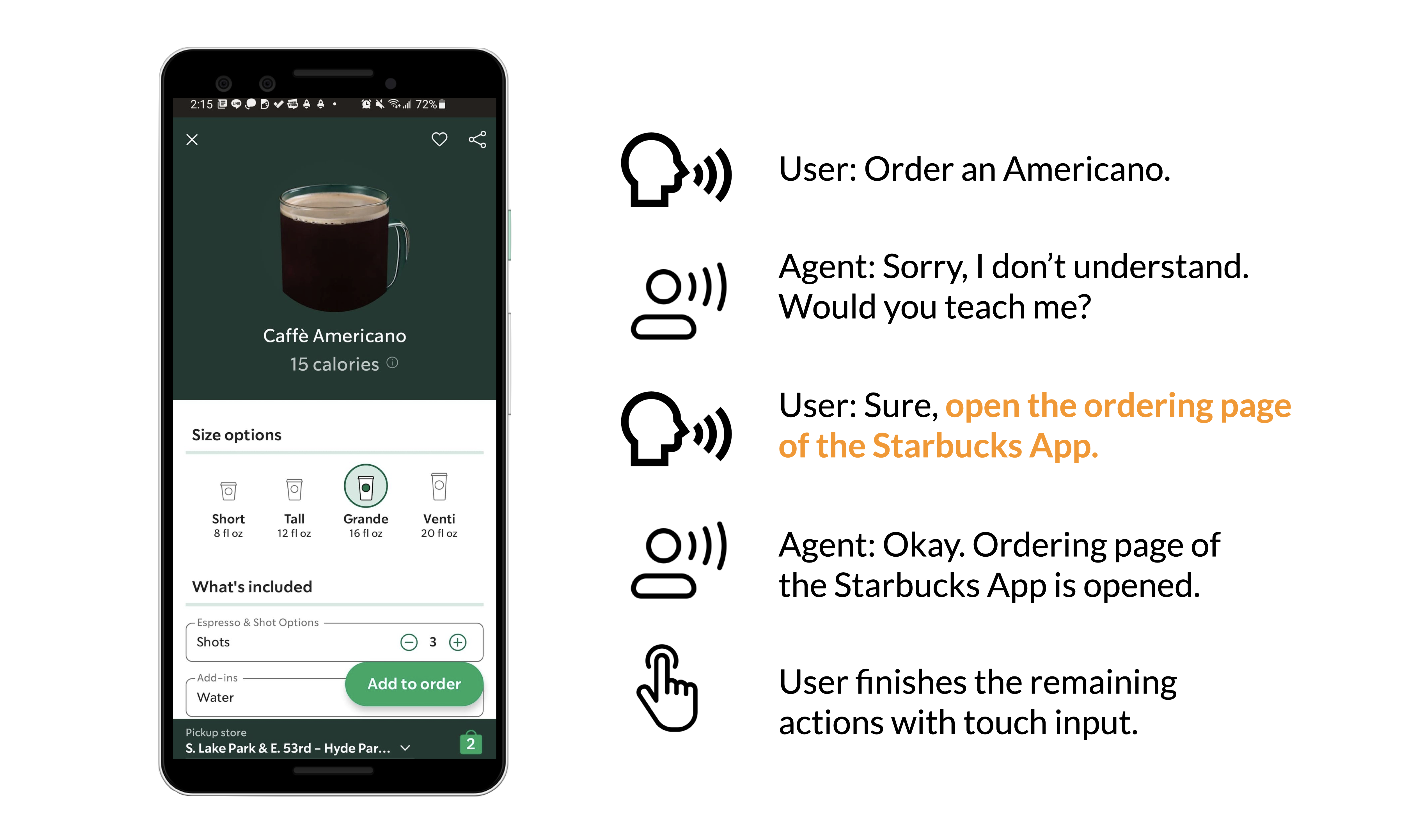}
  \caption{Illustration of how  screen indexing enabled by Screen2Words can facilitate multi-modal interactions to accomplish mobile tasks with conversational agents.}
  \Description{mock up 3}
  \label{fig:mock-up-3}
\end{figure}


\section{Discussions and Future Work}
Screen2Words represents a step closer to bridge mobile UIs with natural language. \bw{The proposed approach is applicable to other tasks that require a holistic understanding of the mobile UI and may generalize to different types of user interfaces such as the web UI.} However, there are still several limitations which future work could address. Firstly, our screen summarization model is not always accurate. As shown in Figure~\ref{fig:model_results}, it sometimes produces generic or wrong summaries. This is akin to the long-standing problem for image captioning where the model is prone to "hallucinating" objects that are not actually in a scene \cite{rohrbach2019object}. \bw{Moreover, similar to prior work on unconventional vision-to-language tasks \cite{wang2018metrics, vs2016huang}, we found that human rating and automatic scores may not always be well-correlated, signaling the need for better automatic evaluation metrics. Future work could investigate new model architectures and evaluation metrics to obtain better summarization results.}

\bw{In some scenarios, app metadata may be missing and the performance of Screen2Words may decrease. However, our approach is still useful as our configurations such as PixelOnly rely on pixel input only---that is always available. Particularly, we found pixel input to be crucial when ScreenText is sparse or entirely absent. 
Such scenarios reinforce the motivation of our multimodal approach, which performs even if a modality is missing.}

Another limitation is that screen summarization only generates descriptions for the entire screen. It cannot be steered by the user to describe the information of a specified section on the screen. Therefore, a natural extension to Screen2Words is towards Visual Question Answering (VQA) \cite{vqa2015, balanced_vqa_v2} for UI screens, which takes a mobile screen and a free-form, open-ended, natural language question as input, and produces a natural language answer as the output. Screen2Words could be viewed as a model specialized for answering questions such as "what's on the screen?". A full-range Screen VQA should be able to answer further questions such as "what actions can I take with the screen?" or "what's the title of the first news article on the screen?" Such technologies could greatly facilitate eye-free, speech interaction with mobile devices.
 
Lastly, one of the use cases of Screen2Words is to facilitate UI retrieval with language queries. An immediate next step is to explore language-based UI generation based on our dataset, which generates mobile UI designs based on language descriptions. Future work could leverage graphical layout generation models \cite{li2019layoutgan, lee2020neural} conditioning on a screen summary to generate user interface designs. To this end, our open-source dataset would be a valuable source to fuel all these research directions.

\section{Conclusion}
We have presented Screen2Words, an approach for automatically summarizing the multi-modal information of a mobile UI screen as a concise language summary using deep learning methods. We collected and analyzed the first large-scale human-annotated dataset to investigate the task. Based on the dataset, we trained and evaluated a set of deep models to examine the feasibility of automatic UI screen summarization. Our evaluation with various automatic metrics shows that deep learning models outperform the heuristic baselines with a significant margin. Our full model, which leverages image, text, and UI structural information, achieves the best results among all the model variants. \bw{We have also conducted a human evaluation using Mechanical Turk and show that our full model significantly outperforms other models and baselines on subjective rating.}
Lastly, we outline three application scenarios that would benefit from the Screen2Words approach. Our dataset, benchmark models and experimental results lay the groundwork for future work on automatic UI screen summarization, which contributes to the effort for bridging natural language and mobile user interface.

\begin{acks}
\bw{We thank the anonymous reviewers for their constructive feedback, which has helped improve this paper. We also thank the participants in our human evaluation and Google data team for their tremendous help on data collection.}
\end{acks}

\bibliographystyle{ACM-Reference-Format}
\bibliography{Screen2Words}



\end{document}